\documentclass[journal,twoside,web]{ieeecolor}
\usepackage{jsen1}
\usepackage{cite}
\usepackage{amsmath,amssymb,amsfonts}
\usepackage{algorithmic}
\usepackage{graphicx}
\usepackage{textcomp}
\usepackage{wrapfig}
\usepackage{subfig}
\usepackage{array}
\def\BibTeX{{\rm B\kern-.05em{\sc i\kern-.025em b}\kern-.08em
    T\kern-.1667em\lower.7ex\hbox{E}\kern-.125emX}}
\definecolor{abstractbg}{rgb}{0.89804,0.94510,0.83137}
\begin{document}
\title{Simultaneous Localization and Recognition of Subwavelength Non-Cooperative Entities Based on SISO Time Reversal and Neural Networks}
\author{ Yinchen Wang, Yu Duan, Yuqi Ye, Ren Wang, \IEEEmembership{Member, IEEE}, Biao Li, Bin Jiang, Xin Liu, and Bingzhong Wang, \IEEEmembership{Senior Member, IEEE}
\thanks{Manuscript received April 7, 2024. This work was supported in part by the National Natural Science Foundation of China under Grants 62171081 and U2341207, the Natural Science Foundation of Sichuan Province under Grant 2022NSFSC0039, and the Aeronautical Science Foundation of China under Grant 2023Z062080002. (Yinchen Wang, Yu Duan, and Yuqi Ye contribute equally to the paper. Corresponding author: Ren Wang.)}
\thanks{Yinchen Wang, Yu Duan, Yuqi Ye, Ren Wang, Biao Li, Xin Liu, and Bingzhong Wang are with the Institute of Applied Physics, University of Electronic Science and Technology of China, Chengdu 611731, China. (e-mail: rwang@uestc.edu.cn).}
\thanks{Bin Jiang is with the Communication Information Security Control Laboratory, No. 36 Research Institute of CETC, Jiaxing 314033, China.}}

\IEEEtitleabstractindextext{%
\fcolorbox{abstractbg}{abstractbg}{%
\begin{minipage}{\textwidth}%
\begin{wrapfigure}[18]{r}{3.7in}%
\includegraphics[width=3.5in]{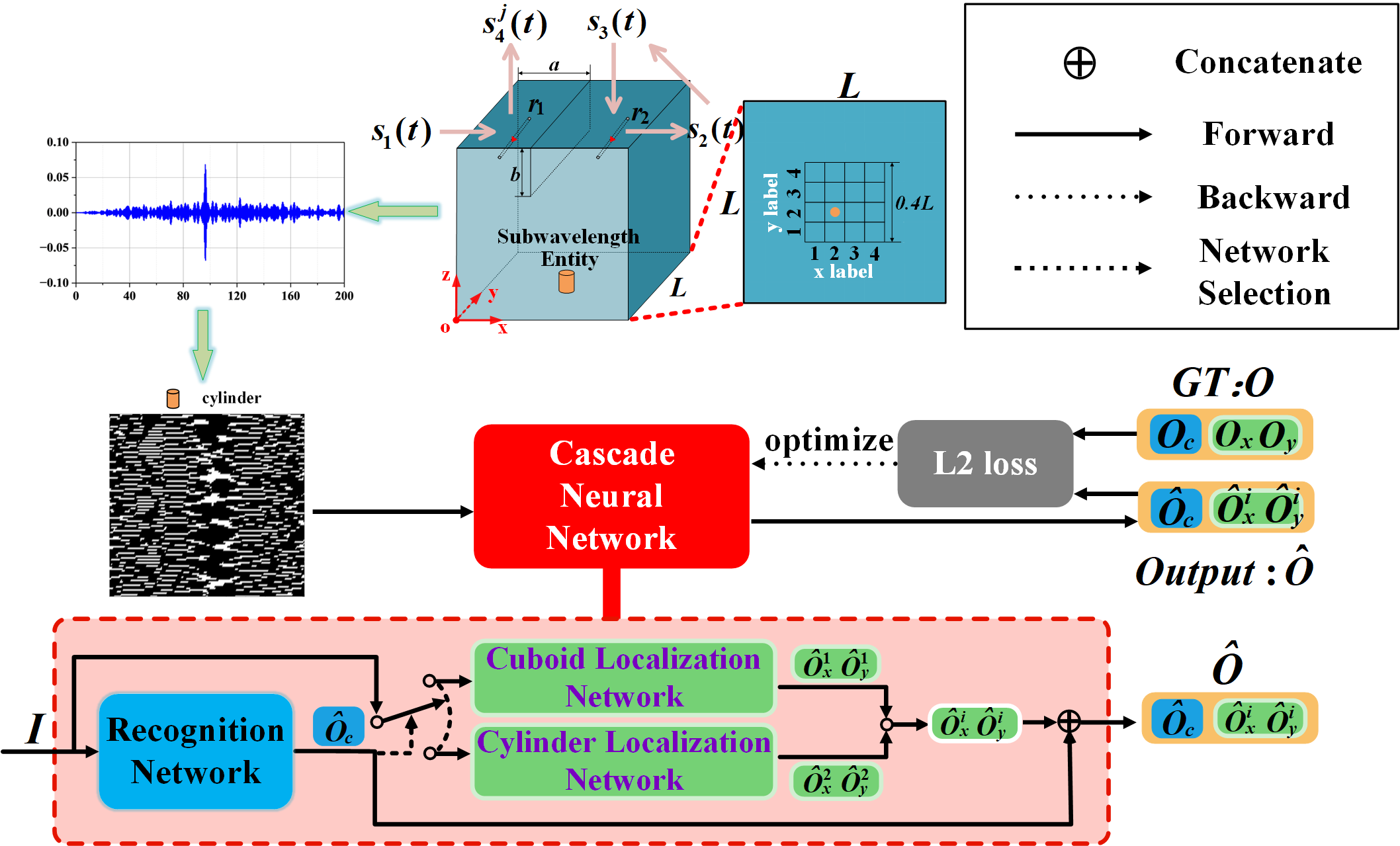}%
\end{wrapfigure}%
\begin{abstract}
The simultaneous localization and recognition of subwavelength non-cooperative entities within complex multi-scattering environments using a simplified system continues to pose a substantial challenge.
In this paper, we address this challenge by synergistically integrating time reversal time-frequency phase prints (TRTFPPs) and neural networks.
Initially, a time reversal (TR) single-input single-output (SISO) framework is employed to generate TRTFPPs.
To enhance the models' adaptability, particularly in the presence of noise, data augmentation techniques are applied.
Subsequently, neural networks are employed to comprehend the TRTFPPs.
Specifically, a cascaded neural network structure is embraced, encompassing both a recognition neural network and distinct neural networks for localizing different entities.
Through the devised approach, two types of subwavelength entities are successfully identified and precisely localized through numerical simulations and experimental verification in laboratory environment.
The proposed methodology holds applicability across various electromagnetic systems, including but not limited to detection, imaging, human-computer interaction, and the Internet of Things (IoT).
\end{abstract}

\end{minipage}}}

\maketitle

\section{Introduction}
\label{sec:introduction}
\IEEEPARstart{N}{ON}-cooperative localization and recognition within multi-scattering scenarios hold paramount significance across domains like detection, imaging, human-computer interaction, and the Internet of Things (IoT). Various methodologies, machine learning, fingerprinting and channel state information analysis, have been proposed to address non-cooperative localization and recognition \cite{b1,b2,b3,b4,b5,b6,b7,b8,b9,b10}. Particularly noteworthy is the accomplishment of simultaneous non-cooperative localization and recognition, achieved through the utilization of single-input single-output (SISO) radar, wavelet features, multiple-input–multiple-output (MIMO) radar, and ultrawideband (UWB) radar in \cite{b11,b12,b13,b14} respectively. However, the resolution limits in these instances lead to the loss of subwavelength information.

In 2021, a remarkable breakthrough in subwavelength localization was attained in a multi-scattering reverberation cavity, employing a SISO system assisted by programmable metasurfaces \cite{b15}. Notably, while a straightforward SISO system is employed, the inclusion of metasurfaces remains imperative for achieving non-cooperative subwavelength localization within multi-scattering scenarios, thereby adding complexity. Furthermore, the endeavor in \cite{b15} did not encompass recognition.

Recent years have seen the burgeoning interest in the time reversal (TR) technique, largely due to its subwavelength focusing capabilities \cite{b16,b17}. Within the past few years, a compelling and uncomplicated SISO subwavelength localization technique under multi-scattering conditions was proposed, leveraging TR fingerprints \cite{b18,b19}. However, these initiatives solely focused on the localization of cooperative sources. Hence, the persistent challenge lies in concurrently achieving subwavelength non-cooperative localization and recognition within multi-scattering contexts.

This paper presents a novel method to achieve simultaneous localization and recognition of subwavelength non-cooperative entities in multi-scattering environments. Within a foundational TR-SISO framework, this method creatively integrates time reversal time-frequency phase prints (TRTFPPs) with a recognition and localization cascade neural network (RLCNN), effectively exploiting the exceptional performance of TRTFPPs through a minimalistic neural network architecture. The proposed scheme exhibits proficiency even in handling deeply sub-wavelength entities, as demonstrated through the successful localization of objects at 1/32 wavelength. The versatility of this method extends to various electromagnetic systems including detection, imaging, human-computer interaction, and the IoT, among others.

\section{Simultaneous Localization and Recognition Method}
\label{sectionII}

\subsection{Calculation of The TRTFPPs}
\begin{figure}[!t]
    \centerline{\includegraphics[width=\columnwidth]{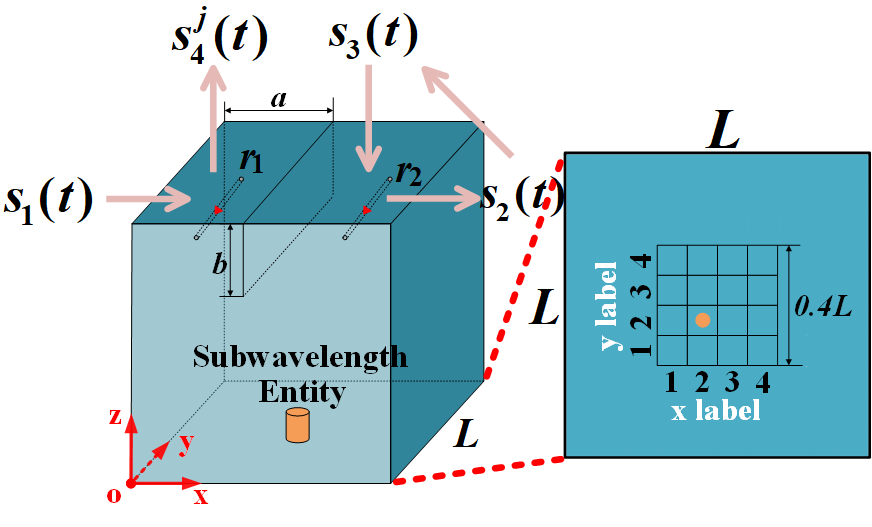}}
    \caption{Scenario of the simultaneous subwavelength localization and recognition method.}
    \label{Fig1}
\end{figure}
Fig. \ref{Fig1} depicts the scenario of proposed methodology. The simulation framework comprises four core components: a cubic metallic cavity with a side length (\emph{L}) of 60 cm to emulate a complex multi-scattering environment, two dipole antennas (positioned at ${r_{1} = (147,295,500)}$ mm and ${r_{2} = (449.5,305,500)}$ mm, respectively) operating at a central frequency of 2.5 GHz, a metallic rectangular clapboard with dimensions of ${L = 60}$ cm and ${b = 15}$ cm, designed to eliminate direct wave propagation between the transmit and receive antennas, and a subwavelength object situated on the ground surface. Notably, for the sake of asymmetry, the clapboard is intentionally positioned away from the cavity’s center, with a specific offset denoted as ‘\emph{a}’ (${a = 25}$ cm). For clarity, a cylinder and a cuboid are chosen to exemplify the subwavelength objects. The dimensions of the cuboid are 12 mm ${\times}$ 12 mm ${\times}$ 30 mm, while the cylinder has a diameter of 12 mm and a height of 30 mm. Both configurations ensure that their largest dimensions, in this case, 0.25 wavelengths at 2.5 GHz, remain beneath the diffraction limit. Furthermore, the ground area’s central region is partitioned into a 4 ${\times}$ 4 grid, with each grid unit having a side length of 60 mm (0.5 wavelengths), collectively spanning an area of 0.4\emph{L} ${\times}$ 0.4\emph{L}. As elaborated subsequently, distinct TRTFPPs emerge from the presence of different subwavelength objects at the same position or the same subwavelength object at varying locations, forming the crux of subwavelength object recognition and localization.

\begin{figure}[!t]
    \centering  
    \subfloat[]	{\includegraphics[width=0.6\columnwidth]{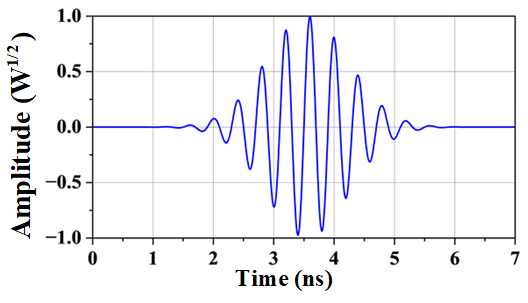}}\\
    \subfloat[]	{\includegraphics[width=0.6\columnwidth]{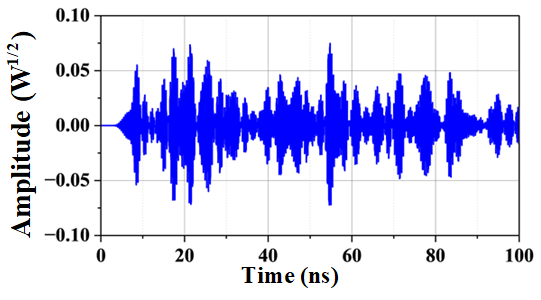}}\\
    \subfloat[]	{\includegraphics[width=0.6\columnwidth]{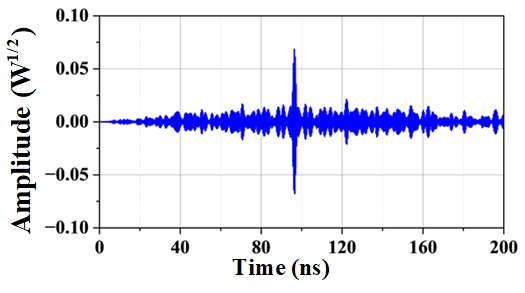}}\\
    \subfloat[]	{\includegraphics[width=0.6\columnwidth]{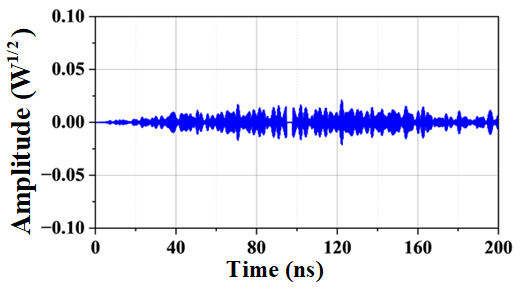}}
    \caption{Signals in the processes: (a) ${s_{1} (t)}$, (b) ${s_{2} (t)}$, (c) ${s_{4} (t)}$, and (d) ${s_{5} (t)}$.}
    \label{Fig2}
\end{figure}
\begin{figure}[!t]
    \centering  
    \subfloat[]	{\includegraphics[width=0.5\columnwidth]{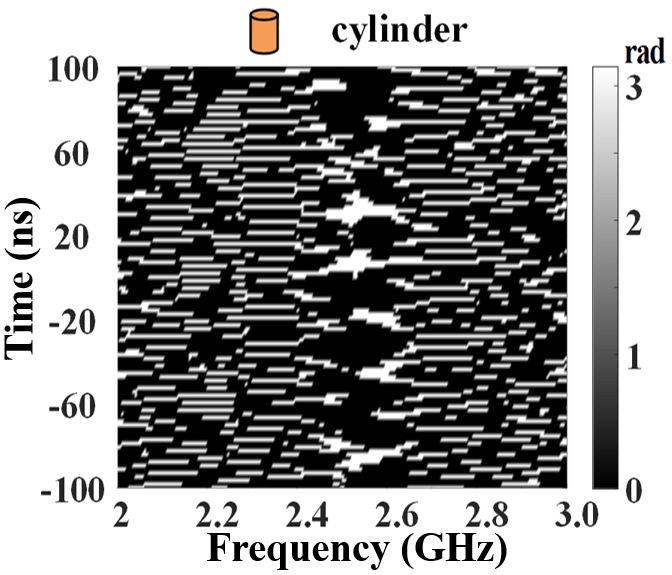}}
    \subfloat[]	{\includegraphics[width=0.5\columnwidth]{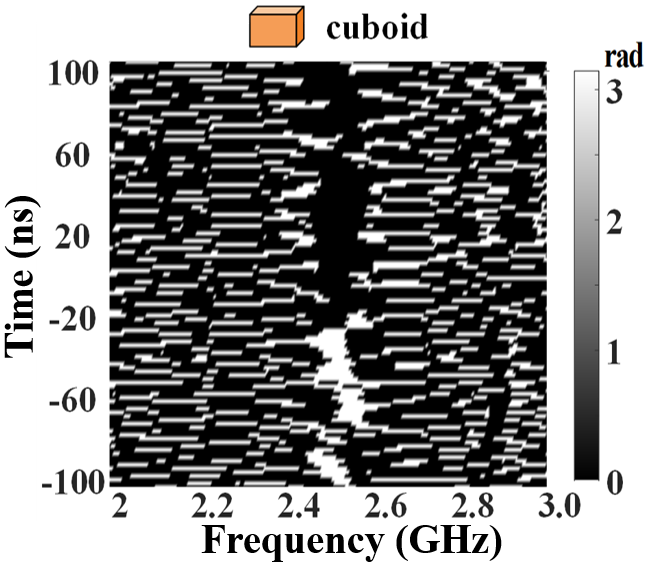}}
    \caption{TRTFPPs corresponding to (a) the cuboid and (b) the cylinder in the grid (1,1).}
    \label{Fig3}
\end{figure}

In the environment detection process, a 2-3 GHz Gaussian signal $s_{1}\left ( t \right )$ (Fig. \ref{Fig2}(a)) is transmitted from the antenna ${r_{1}}$ into the cavity when there is no subwavelength objects. Then the antenna ${r_{2}}$ receives a signal ${s_{2} (t) = s_{1} (t)\otimes h_{0} (t)}$ (Fig. \ref{Fig2}(b)), where ${h_{0} (t)}$ is the channel response function of the empty cavity and ${\otimes }$ represents convolution. Next, the signal ${s_{2} (t)}$ is time-reversed to get ${s_{3}(t) = s_{2}(-t) = s_{1}(-t) \otimes h_{0}(-t)}$. In the TRTFPP generation, the time-reversed signal ${s_{3} (t)}$ is transmitted from the antenna ${r_{2}}$ into the cavity when there is a subwavelength object in the grid. Then the antenna ${r_{1}}$ receives the signal ${s_{4}^{i}(t) = s_{3}(t) \otimes h_{i}(t) = s_{1}(-t) \otimes h_{0}(-t) \otimes h_{i}(t)}$ (Fig. \ref{Fig2}(c)), where ${h_{i}(t)}$ is the channel response function depending on the subwavelength object and its position. The TR time-frequency print ${E_{i}^{TR} (t,\omega )}$ can be calculated with the method of discrete short-time fourier transform (DSTFT) \cite{b20,b21,b22}:
\begin{equation}
    \begin{aligned}
    E_{i}^{TR}(t,\omega ) \approx \int\limits_{-\infty }^{+\infty } \left [ s_{1}(-\tau ) \otimes h_{0}(-\tau ) \otimes h_{i}(\tau )  \right ] 
    \\ \cdot R(\tau ) \cdot W(\tau -t) e^{-j\omega \tau } d\tau 
    \end{aligned}
\end{equation}

\noindent 
where ${W(\tau -t)}$ is the rectangular window function with the width of 16 ns and ${R(\tau )}$ denotes the removal operation of the TR focus peak, and the removed temporal width is related to the bandwidth of ${s_{1}(-\tau )}$. In Fig. \ref{Fig2}(d), ${s_{5}^{i}(t) = s_{4}^{i}(t) \cdot R(t)}$. The phase of the TR time-frequency print ${Arg\left [ E_{i}^{TR}(t,\omega )  \right ]}$ is the TRTFPP. The TRTFPPs corresponding to the cuboid and the cylinder in the grid (1,1) are shown in Fig. \ref{Fig3}, clearly demonstrating the noticeable difference between them.

\subsection{Data Augmentation}
Following the acquisition of TRTFPPs, the subsequent step involves employing neural network technology for subwavelength object recognition and positional prediction. To effectively simulate a recognition environment marred by noise and bolster the resilience of neural networks, a data augmentation strategy is implemented. This strategy encompasses the introduction of Gaussian noise to all received signals during the neural network training process \cite{b23}:
\begin{align}
    S_{4}^{i}(t) = s_{4}^{i} + n_{i}(t)
\end{align}

\noindent where ${n_(t)}$  represents Gaussian noise with zero-mean and variance ${\sigma ^{2} }$. By setting ${\sigma =}$ 0.0005, 0.0010, 0.0015, and 0.0020, different level of Gaussian noise is added into the received signal. The noisy TRTFPPs can be implemented by replacing ${s_{4}^{i}}$ with ${S_{4}^{i}(t)}$ in the above method. The data augmentation process encompasses the generation of five diverse noisy iterations for each scenario, incorporating the case of ${\sigma = 0}$. This procedure culminates in the production of a total of 160 distinct TRTFPPs, all pertaining to cylinders and cuboids distributed across 16 grids.

\subsection{Settings of Neural Network}
\begin{figure}[!t]
    \centerline{\includegraphics[width=\columnwidth]{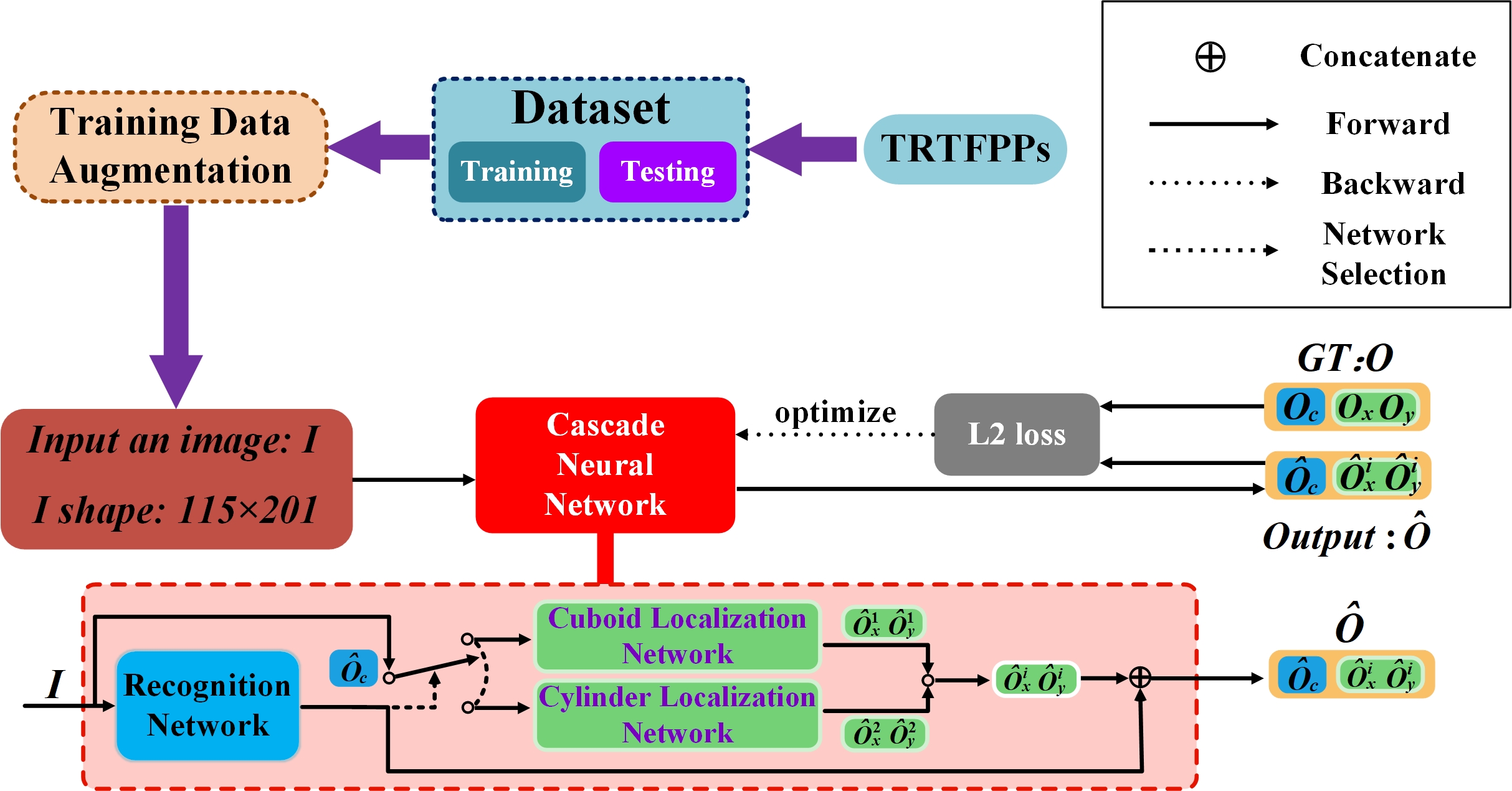}}
    \caption{Recognition and localization cascade neural network.}
    \label{Fig4}
\end{figure}

\begin{figure}[!t]
    \centerline{\includegraphics[width=\columnwidth]{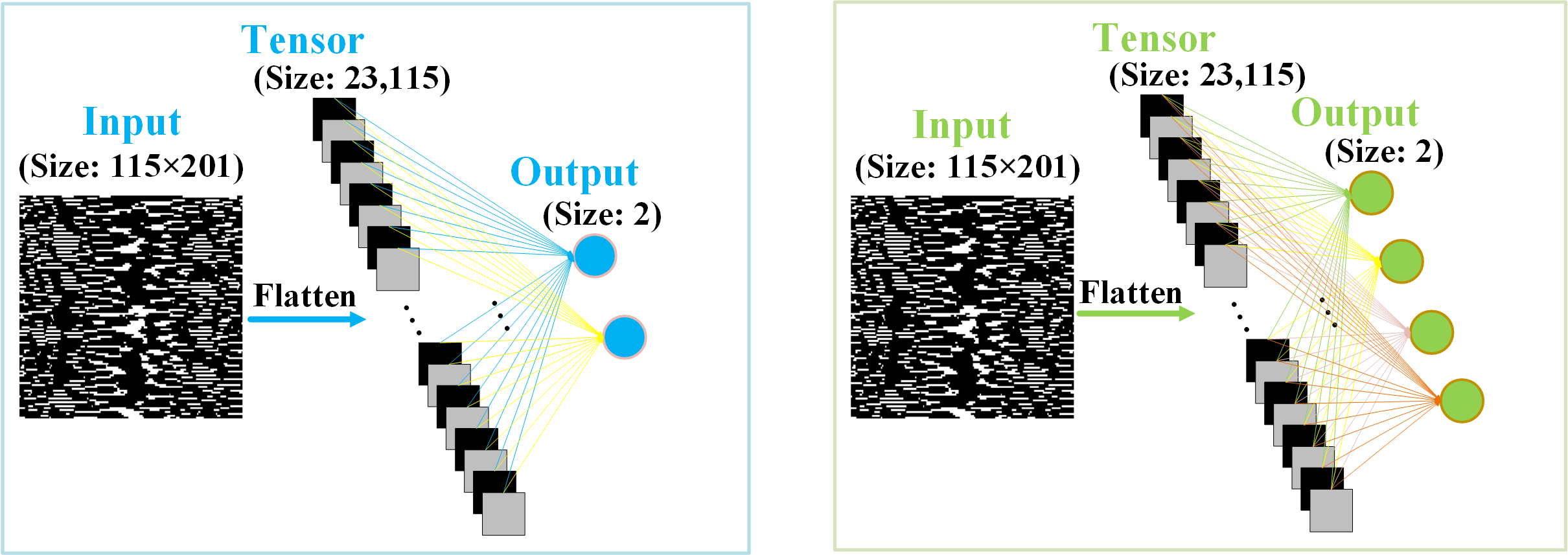}}
    \caption{Neural network architectures: (a) recognition network and (b) localization network.}
    \label{Fig5}
\end{figure}

The TRTFPPs play a pivotal role in the training of the RLCNN. The RLCNN consists of three distinct sub-networks: a recognition network, a cuboid localization network, and a cylinder localization network, as depicted in Fig. \ref{Fig4}. To emphasize the effectiveness of the proposed method, very simple networks are employed, with each sub-network comprising only one fully connected layer, as shown in Fig. \ref{Fig5}. The recognition network determines whether an input TRTFPP belongs to a cuboid or a cylinder. Subsequently, the localization network locates the horizontal label (x) and vertical label (y) through a fully connected layer. The recognition network exhibits dynamic selection of the appropriate localization network based on distinct TRTFPPs. In detail, the recognition network processes the TRTFPP to yield a recognition output denoted as  ${\widehat{O} _{c}}$. Based on  ${\widehat{O} _{c}}$, the TRTFPP is then channelled into the selected localization network, yielding the positional output ${( \widehat{O}_{x}^{i} , \widehat{O}_{y}^{i}) }$. To optimize the neural network parameters and mitigate overfitting, the regularized quadratic loss function (L2 loss) is employed, reflecting the disparity between the output ${\widehat{O}}$ and the ground truth (GT) ${O}$ \cite{b24}. Leveraging learning rate decay and exponential moving average augments the networks’ robustness during training initiation and ensures improved performance on test datasets \cite{b25,b26}.

Given that the recognition network’s primary function is to differentiate between cylinder and cuboid objects, its training employs 24 out of 32 noise-free TRTFPPs, while the remaining 8 are utilized for testing, randomly chosen. The training sets for each of the two localization networks are derived from cylinders or cuboids positioned at the center of the grid under five distinct levels of noise (${\sigma =}$ 0, 0.0005, 0.0010, 0.0015, and 0.0020). Conversely, the test sets consist of cylinders or cuboids situated adjacent to the grid's center (+0, +0.05mm), free from noise (${\sigma =}$ 0), enabling assessment of effectiveness, robustness, and generalization. Collectively, the training dataset encompasses 80 TRTFPPs, while the test dataset comprises 16 TRTFPPs.

\section{Simulation Verification}
\label{sectionIII}

\subsection{Recognition and Localization of Single Entity}
\begin{figure}[!t]
    \centering  
    \subfloat[]	{\includegraphics[width=\columnwidth]{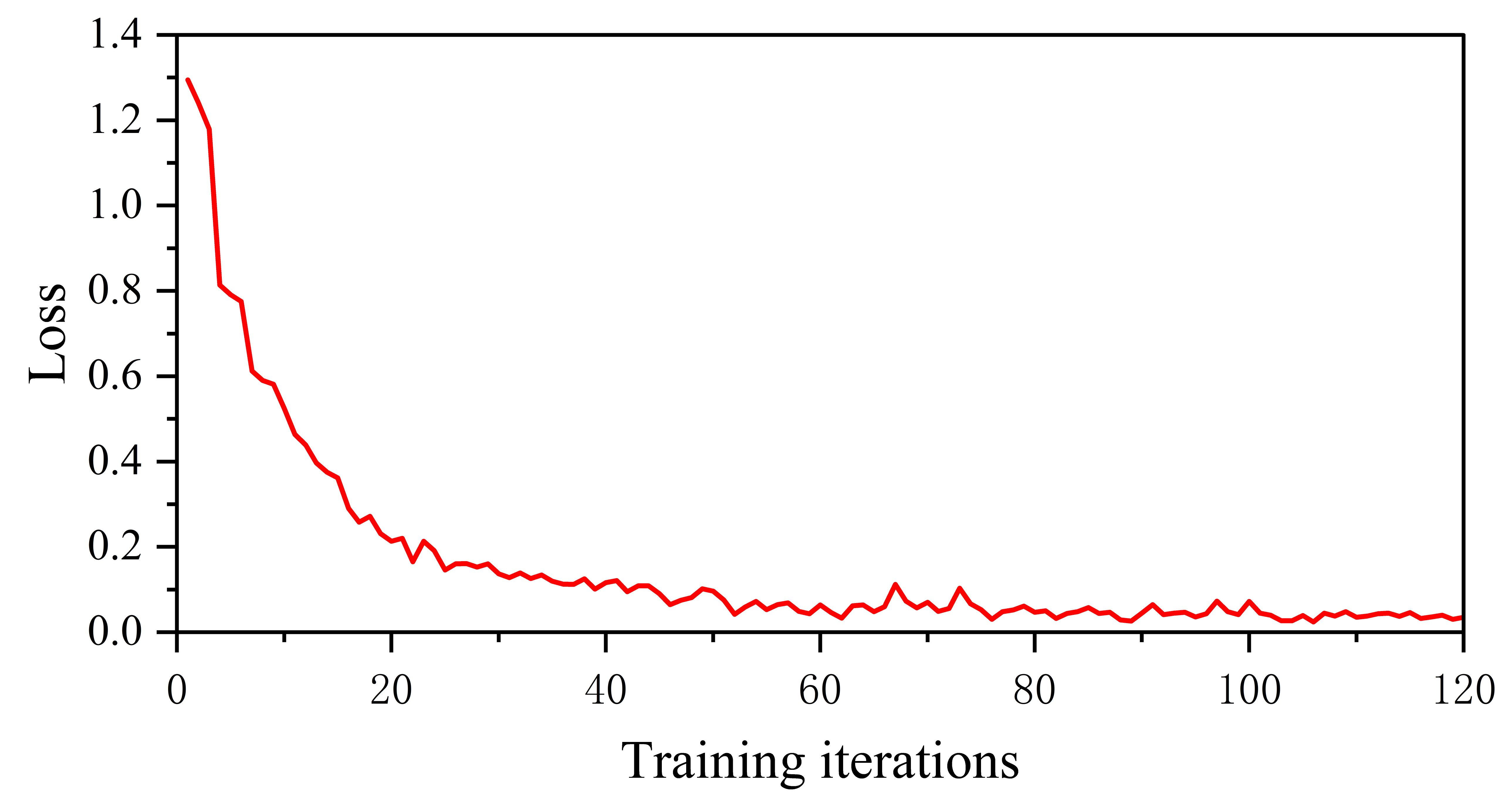}}\\
    \subfloat[]	{\includegraphics[width=\columnwidth]{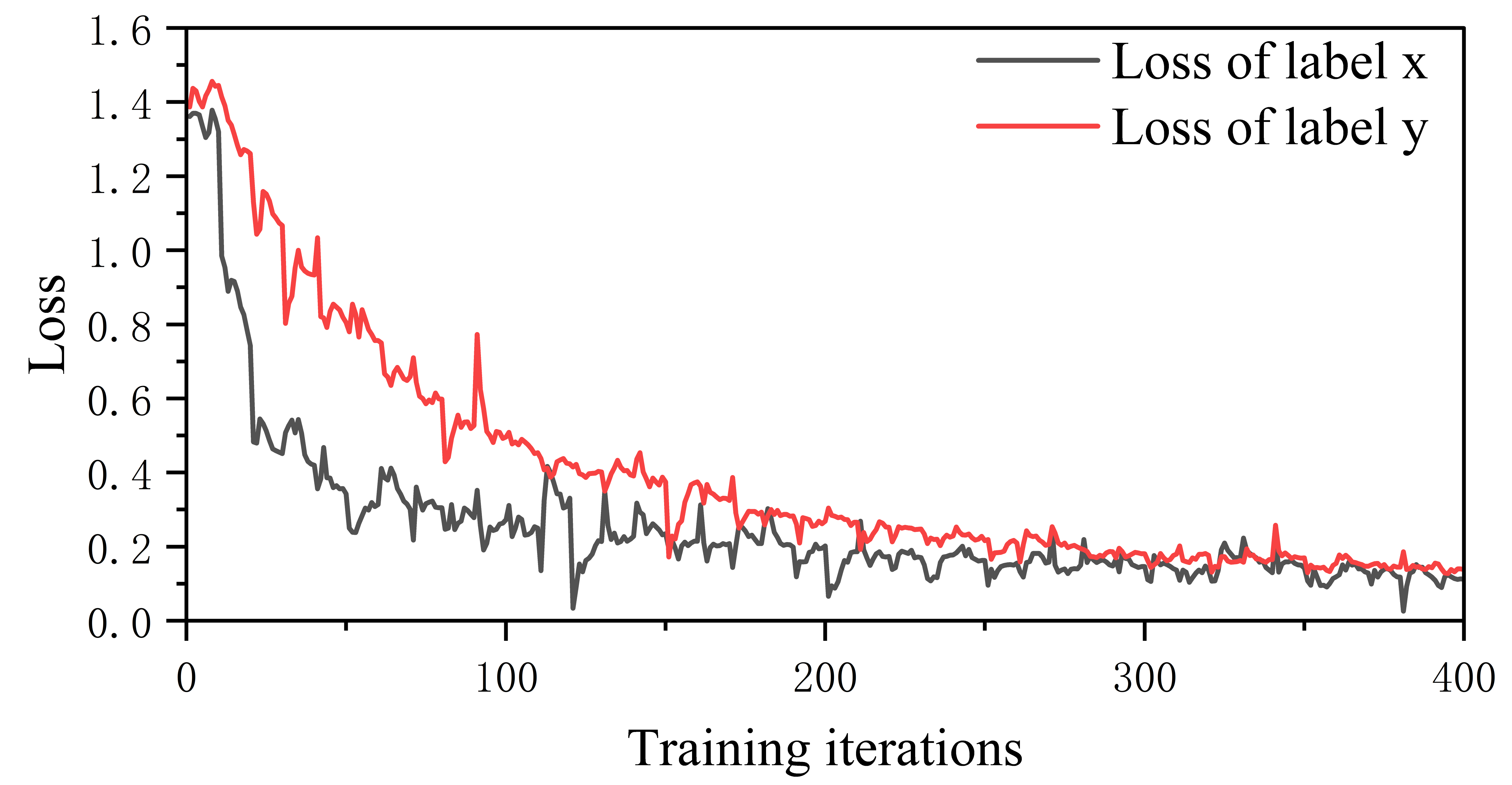}}\\
    \subfloat[]	{\includegraphics[width=\columnwidth]{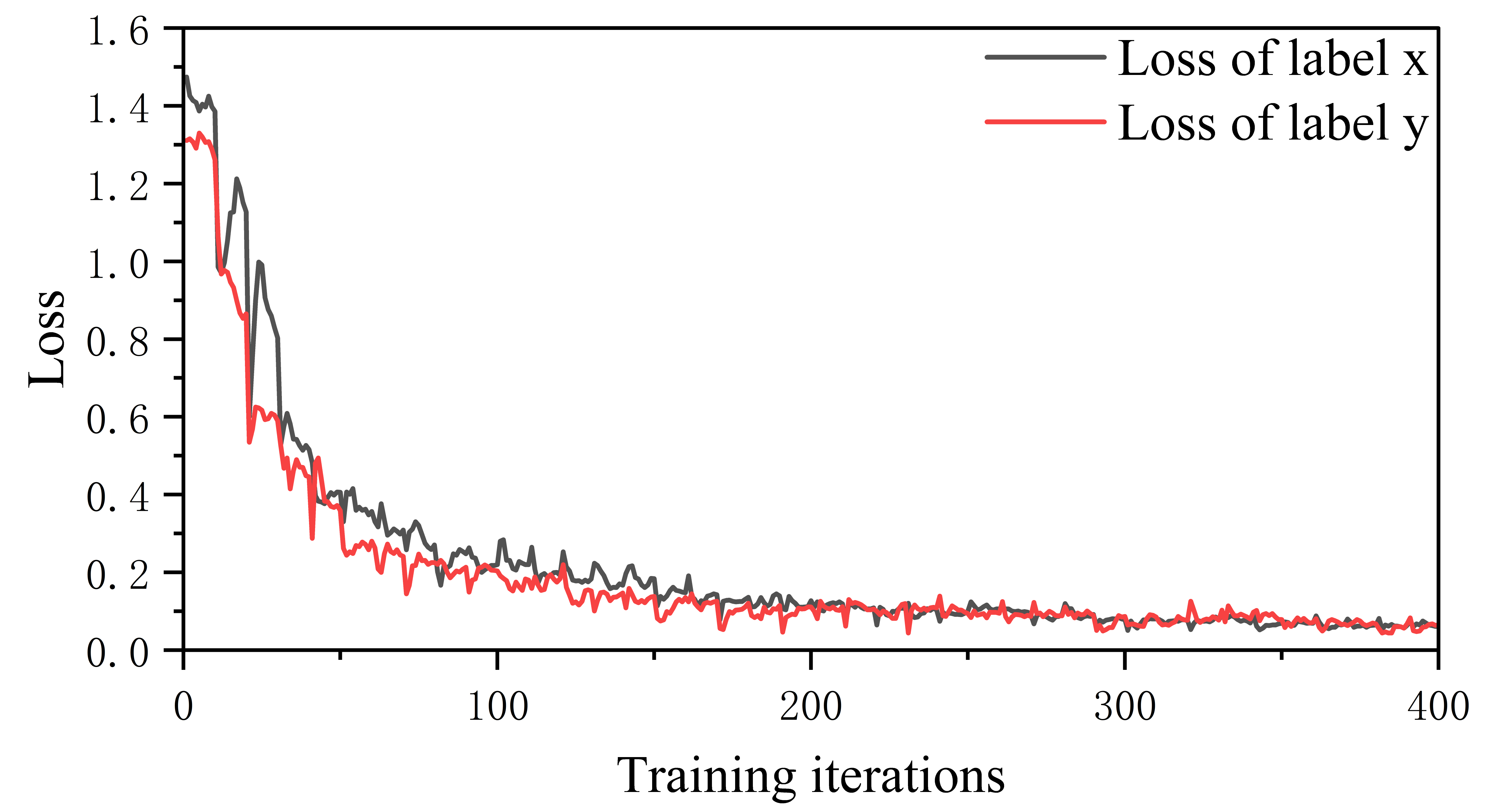}}
    \caption{Training loss of the RLCNN based on simulation data relative to training iteration: (a) recognition, (b) cylinder localization, and (c) cuboid localization.}
    \label{Fig6}
\end{figure}
\begin{figure}[!t]
    \centering  
    \subfloat[]	{\includegraphics[width=0.6\columnwidth]{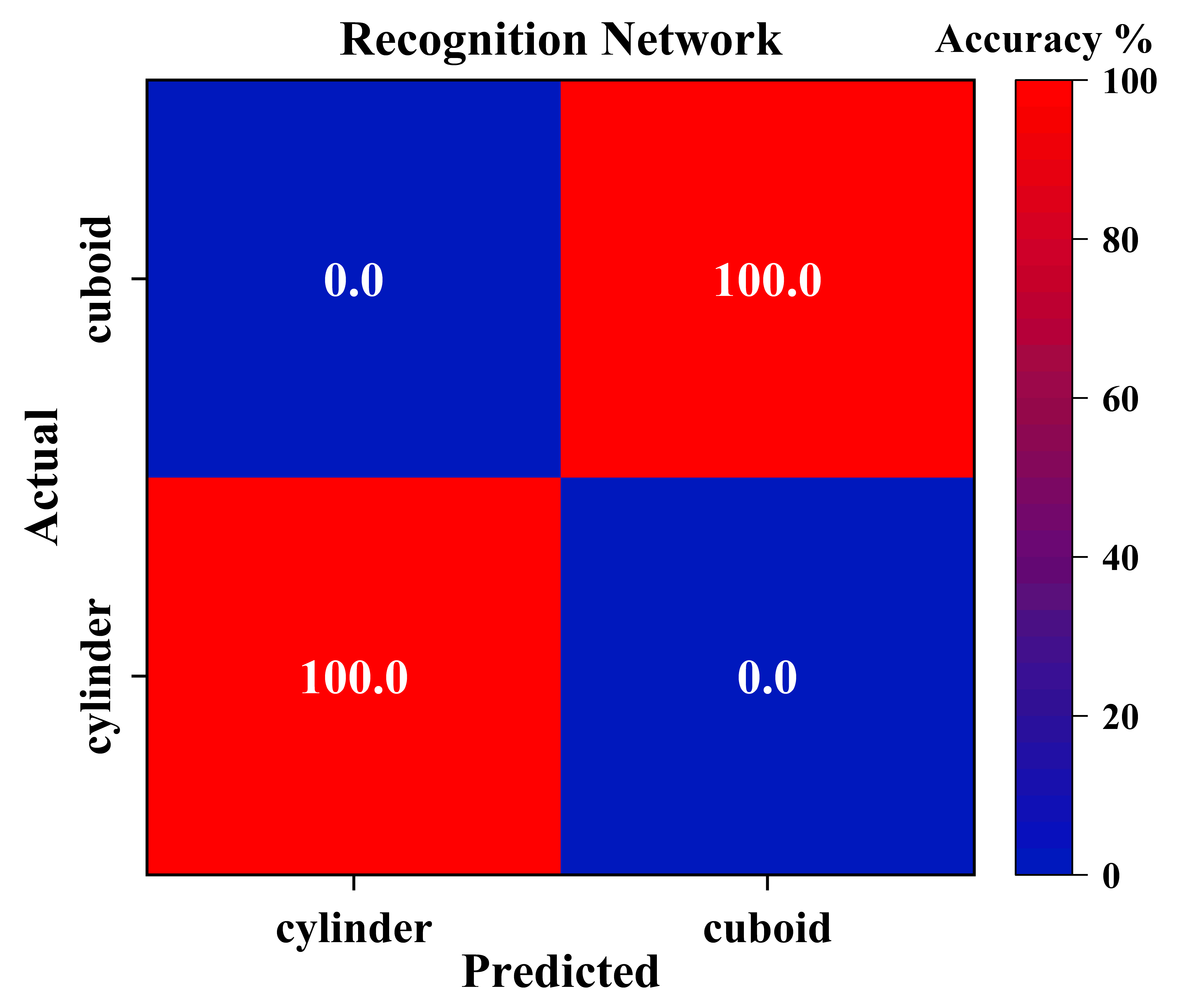}}\\
    \subfloat[]	{\includegraphics[width=\columnwidth]{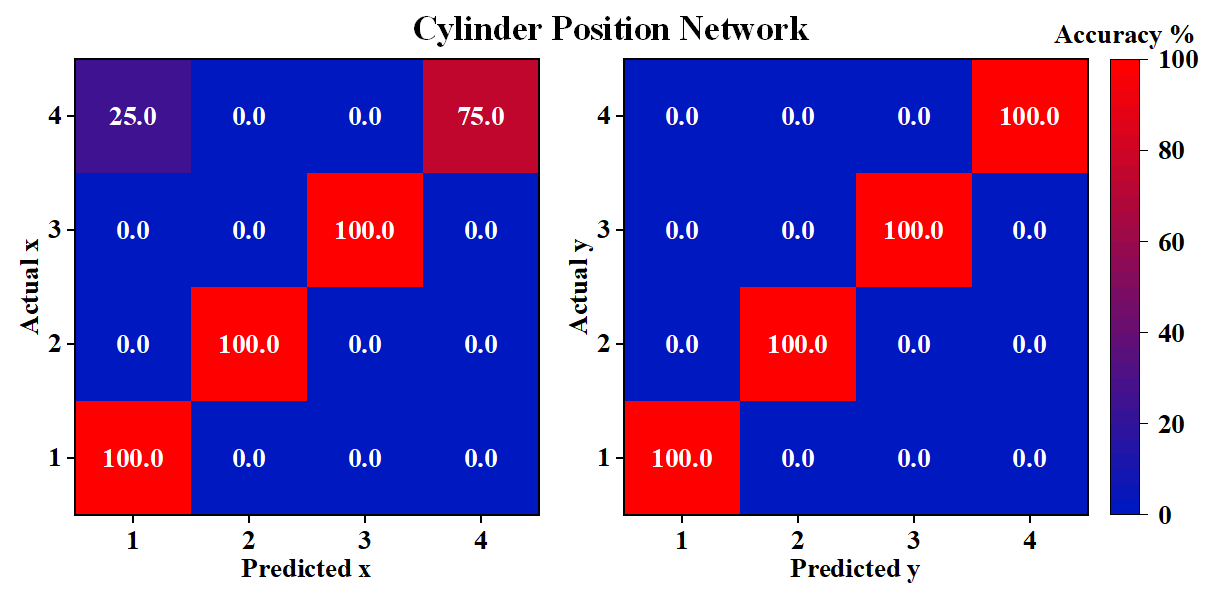}}\\
    \subfloat[]	{\includegraphics[width=\columnwidth]{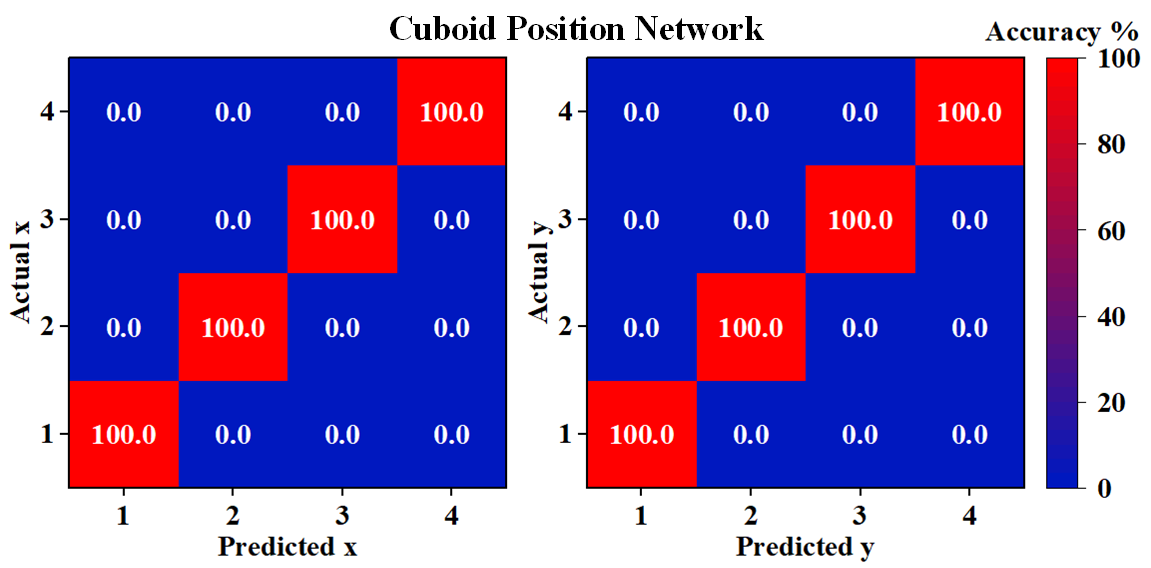}}
    \caption{Heat map of the RLCNN: (a) recognition, (b) cylinder localization, and (c) cuboid localization. The horizontal label (x) and vertical label (y) in (b) and (c) indicate the 16 girds in Fig. \ref{Fig1}.}
    \label{Fig7}
\end{figure}
Figs. \ref{Fig6} and \ref{Fig7} depict the training loss and the heat map generated by the RLCNN, respectively. Evidently, the training loss has successfully converged. Following an intensive training period spanning 120 to 400 iterations, the testing accuracy of the recognition network, cylinder localization network, and cuboid localization network attains remarkable levels of 100\%, 93.75\%, and 100\%, respectively. This exemplary performance reaffirms the RLCNN’s capability to effectively discern subwavelength objects and accurately forecast their precise positions.

\subsection{Higher Resolution Localization}
\begin{figure}[!t]
    \centerline{\includegraphics[width=\columnwidth]{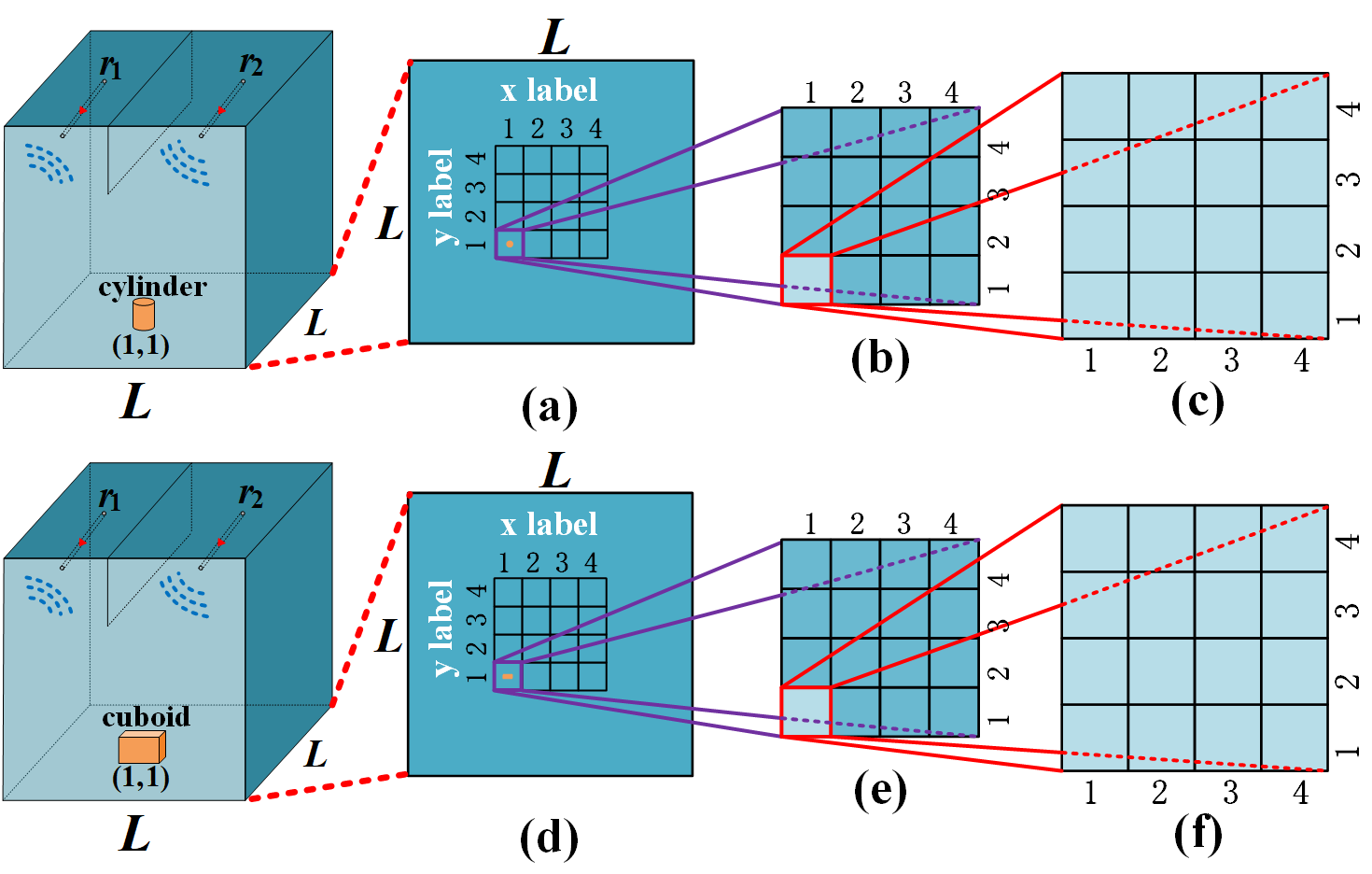}}
    \caption{Schematic illustration of the higher resolution localization experiment: (a)-(c) cylinder, (d)-(f) cuboid. The grid side lengths in (a) and (d), (b) and (e), (c) and (f) are 60 mm (1/2 wavelengths), 15 mm (1/8 wavelengths), and 3.75 mm (1/32 wavelengths), respectively.}
    \label{Fig8}
\end{figure}

Illustrated in Fig. \ref{Fig8}, we extend our investigation by partitioning the aforementioned (1,1) grid into 4 ${\times}$ 4 subdivisions. This approach is then iteratively applied, further dividing the new (1,1) grid into 4 ${\times}$ 4 subdivisions, thereby offering a compelling showcase of the technique’s exceptional super-resolution localization capacity. The grid dimensions in Figs. \ref{Fig8}(a) and \ref{Fig8}(d), Figs. \ref{Fig8}(b) and \ref{Fig8}(e), Figs. \ref{Fig8}(c) and \ref{Fig8}(f) are 60 mm (equivalent to 1/2 wavelengths), 15 mm (equivalent to 1/8 wavelengths), and 3.75 mm (equivalent to 1/32 wavelengths), respectively.

Similar to previous scenarios, the training dataset originates from cylinders or cuboids centered within the grid, incorporating data augmentation techniques. Correspondingly, the test dataset features cylinders or cuboids positioned adjacent to the grid center (+0, +0.05mm). For the localization task, a pair of neural network models containing only one fully connected layer are deployed to estimate the horizontal label (x) and vertical label (y).

\begin{table}[!t]
\renewcommand\arraystretch{1.5}
\centering
\caption{Localization Accuracy of Different \\Resolution Tests}
\label{Table1}
\setlength{\tabcolsep}{10pt}
\begin{tabular}{c c c c}
    \hline
    Different & Accuracy of & Accuracy of & Accuracy of \\
    Resolution & x & y & Position \\ 
    \hline
    Fig. \ref{Fig8}(a) & 93.75\% & 100\% & 93.75\% \\
    Fig. \ref{Fig8}(d) & 100\% & 100\% & 100\% \\
    Fig. \ref{Fig8}(b) & 100\% & 100\% & 100\% \\
    Fig. \ref{Fig8}(e) & 100\% & 100\% & 100\% \\
    Fig. \ref{Fig8}(c) & 87.5\% & 93.75\% & 87.5\% \\
    Fig. \ref{Fig8}(f) & 81.25\% & 93.75\% & 75\% \\
    \hline
\end{tabular}
\end{table}

The outcomes of the testing phase are detailed in Table \ref{Table1}. The data clearly demonstrate that the proposed approach remains effective in accurately predicting object positions, even when dealing with profound subwavelength distances as small as 1/32 wavelengths.

\subsection{Localization and Recognition of Multiple Entities in a More Complex Environment}
\begin{figure}[!t]
    \centerline{\includegraphics[width=\columnwidth]{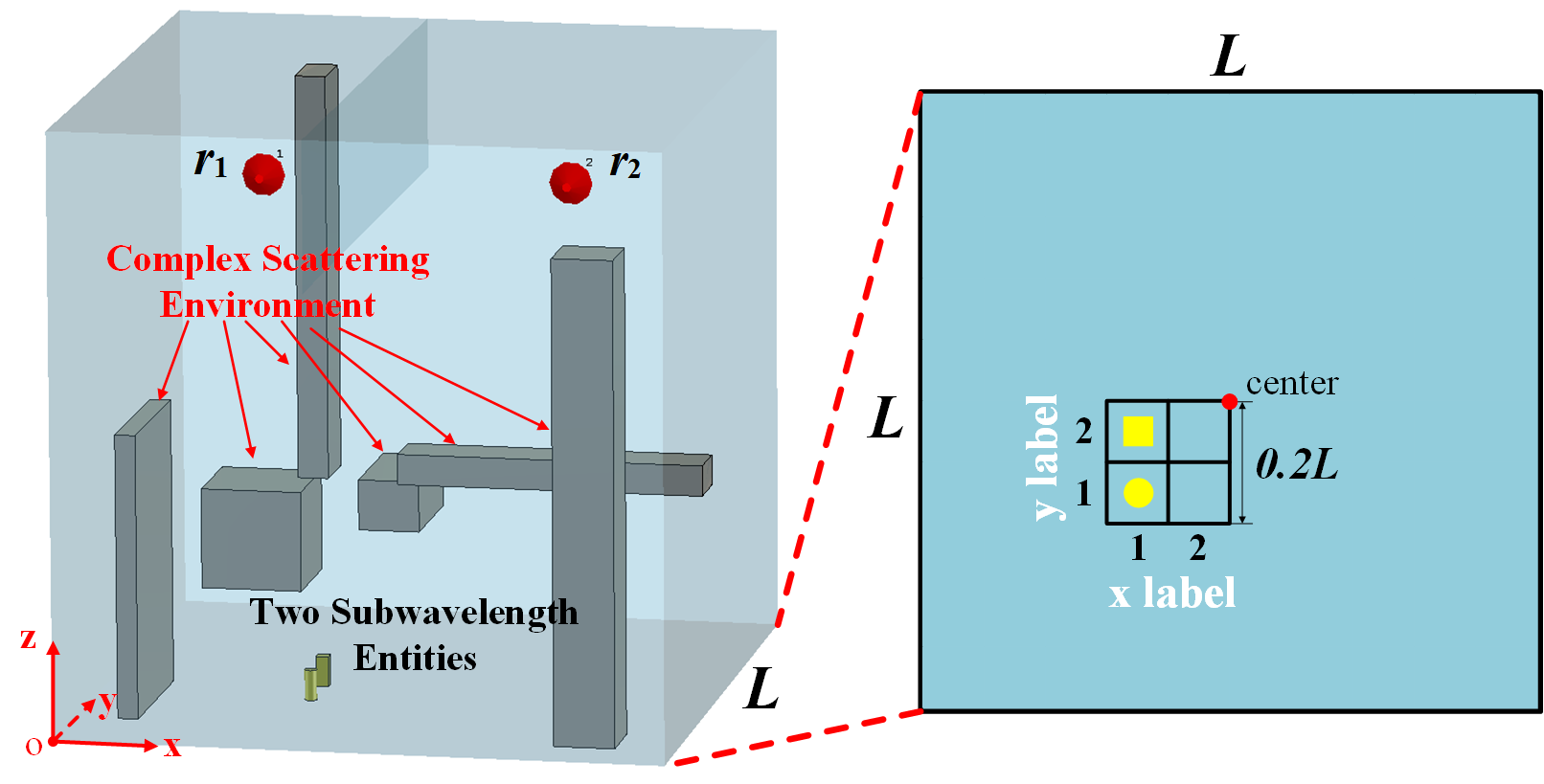}}
    \caption{Schematic illustration of multi-entity recognition and location experiment in complex environment.}
    \label{Fig9}
\end{figure}

In order to demonstrate the robustness of the method to the environment and the scalability of the application, we did a simulation with some random sized metal scatters and two entities to be recognized and localized, as shown in Fig. \ref{Fig9}.

\begin{figure}[!t]
    \centering  
    \subfloat[]	{\includegraphics[width=\columnwidth]{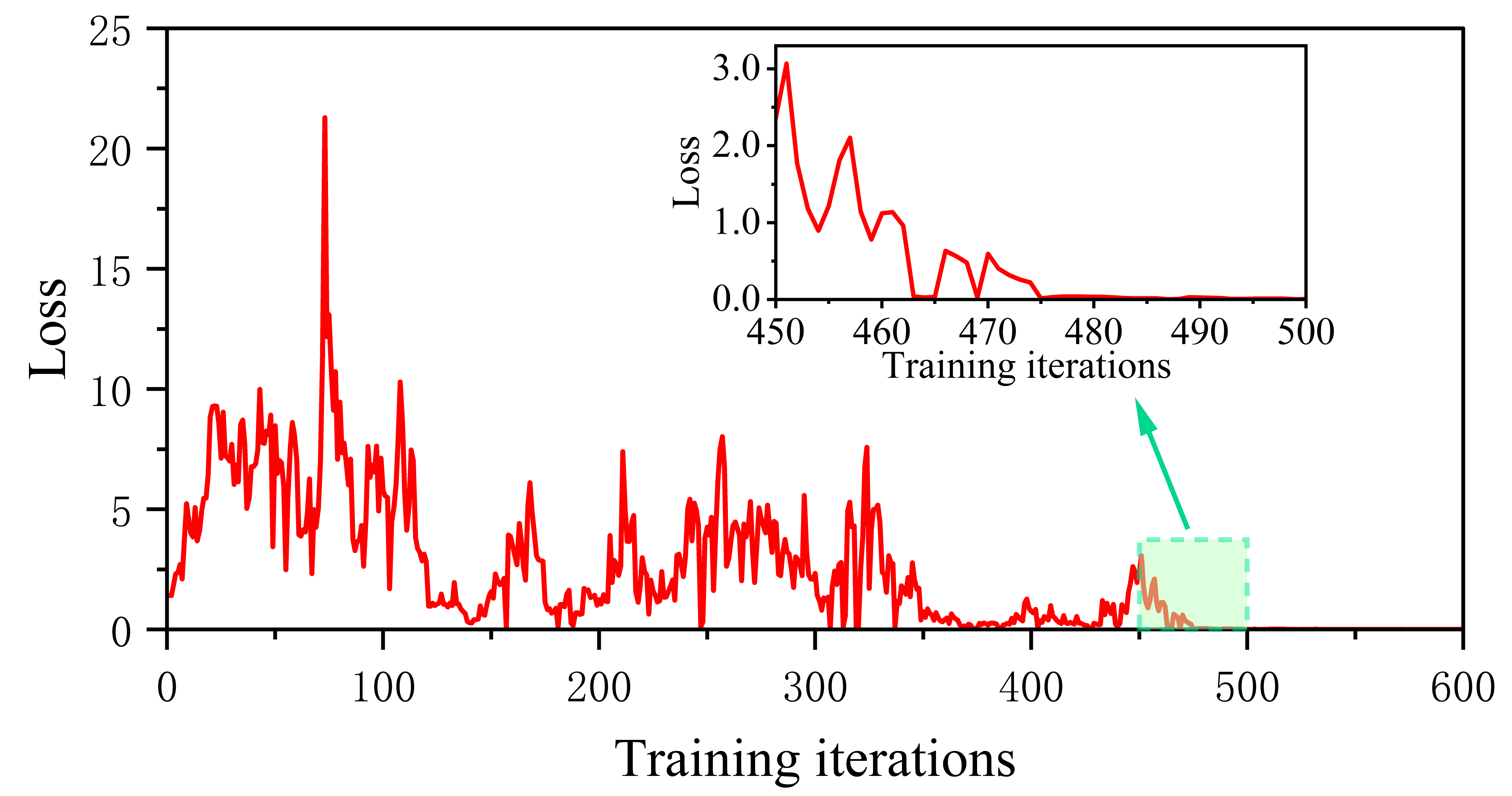}}\\
    \subfloat[]	{\includegraphics[width=\columnwidth]{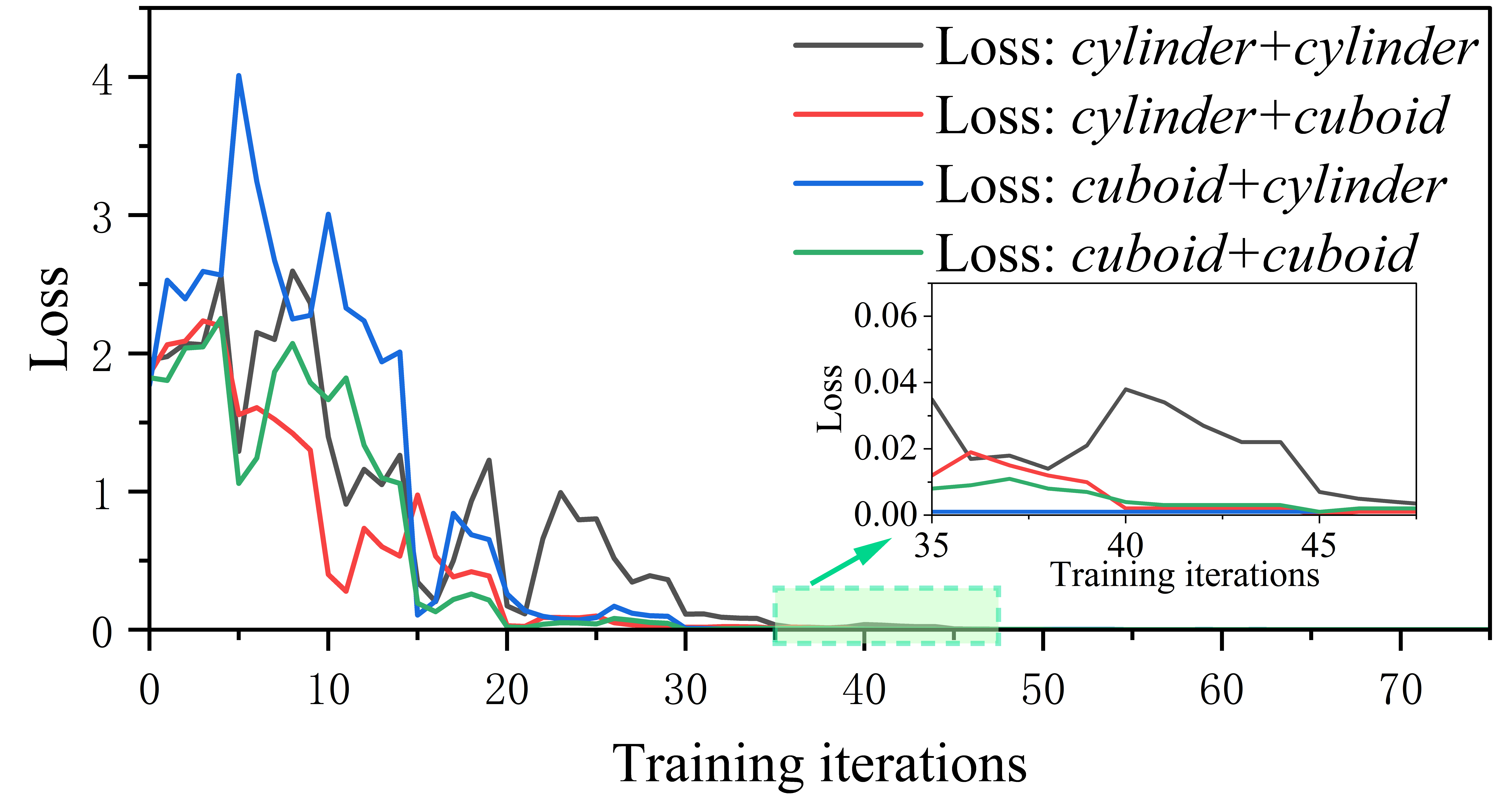}}
    \caption{Training loss of RLCNN in the case of two entities: (a) shape recognition and (b) localization of each shape.}
    \label{Fig10}
\end{figure}

At the bottom of the metal cavity, we divided 2 ${\times}$ 2 grids in the off-center position for positioning, and the side length of each small cube grid was 0.1${L}$, that is, 60mm. And set the horizontal label (x) and the vertical label (y). For the entities identified and located, the dimensions of the cuboid are 12 mm ${\times}$ 12 mm ${\times}$ 30 mm, while the cylinder has a diameter of 12 mm and a height of 30 mm. At the system's operating frequency of 2.5GHz, their maximum size is 1/4 wavelength. When considering placing two entities in 2 ${\times}$ 2 grids simultaneously, there are 24 different ways to place them. Specifically, we define ${cylinder+cylinder}$, ${cylinder+cuboid}$, ${cuboid+cylinder}$, and ${cuboid+cuboid}$ as 4 shapes. According to the method of arrangement and combination, each shape has 6 different position combinations, and we define it as 6 positions.

Based on the method proposed in the paper, we obtain 24 TRTFPPs with entities at the center point of the corresponding grid, and 24 ${\times}$ 4 TRTFPPs through data augmentation techniques (adding Gaussian noise). Finally, a total of 120 TRTFPPs serve as the training dataset for RLCNN. The testing dataset consists of 24 TRTFPPs formed when each entity deviates (+0,+0.05 mm) from the center of the grid.

The training loss and test accuracy of RLCNN are shown in Fig. \ref{Fig10} and Table \ref{Table2}, respectively. It can be seen from the training and testing results that the proposed method still has excellent recognition and localization performance in complex scattering environments, and is very accurate for multi-target recognition and localization. This indicates that the method has a very broad development and application prospects.

\begin{table}[!t]
\renewcommand\arraystretch{1.5}
\centering
\caption{Localization Accuracy of Different Shapes}
\label{Table2}
\setlength{\tabcolsep}{8pt}
\begin{tabular}{c c c c}
    \hline
    Different & Accuracy of & Accuracy of & Accuracy of \\
    Shapes & x & y & Position \\ 
    \hline
    ${cylinder+cylinder}$ & 100\% & 100\% & 100\% \\
    ${cylinder+cuboid}$ & 100\% & 100\% & 100\% \\
    ${cuboid+cylinder}$ & 100\% & 100\% & 100\% \\
    ${cuboid+cuboid}$ & 100\% & 100\% & 100\% \\
    \hline
\end{tabular}
\end{table}

\section{Localization and Recognition Experiment}
The proposed method is validated in the simulation environment outlined in Section \ref{sectionIII}.
In this section, we conduct an experiment to demonstrate the feasibility of the method by recognizing and locating 1/4 wavelength metal scatterers in a complex scattering environment based on the methods and procedures outlined in Section \ref{sectionII}.
\begin{figure}[!t]
    \centerline{\includegraphics[width=\columnwidth]{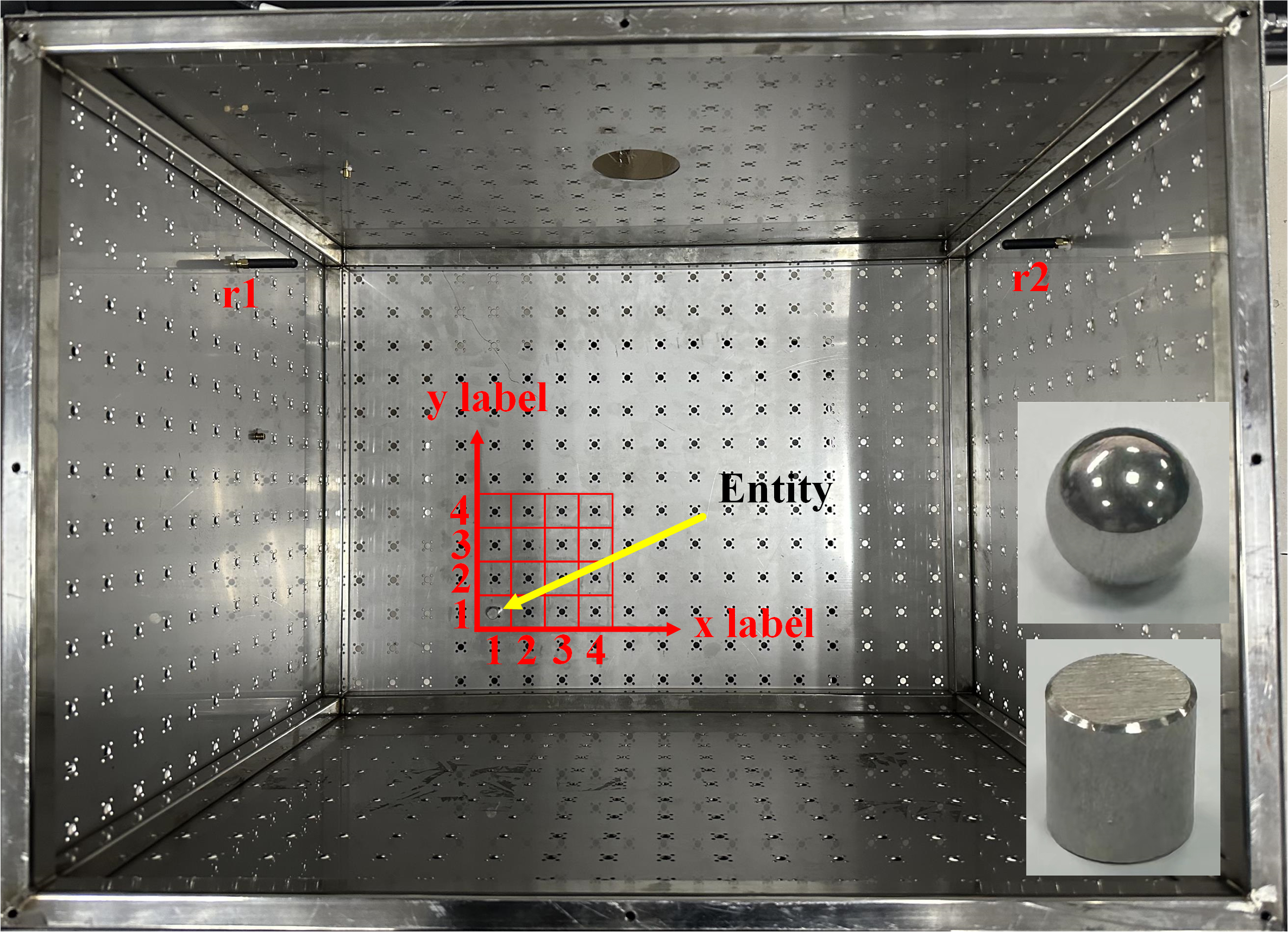}}
    \caption{The interior of the metal cavity and its meshing (top view). The inserted pictures are the sphere and the cylinder for recognition and localization.}
    \label{Fig11}
\end{figure}

\begin{figure}[!t]
    \centering  
    \subfloat[]	{\includegraphics[width=\columnwidth]{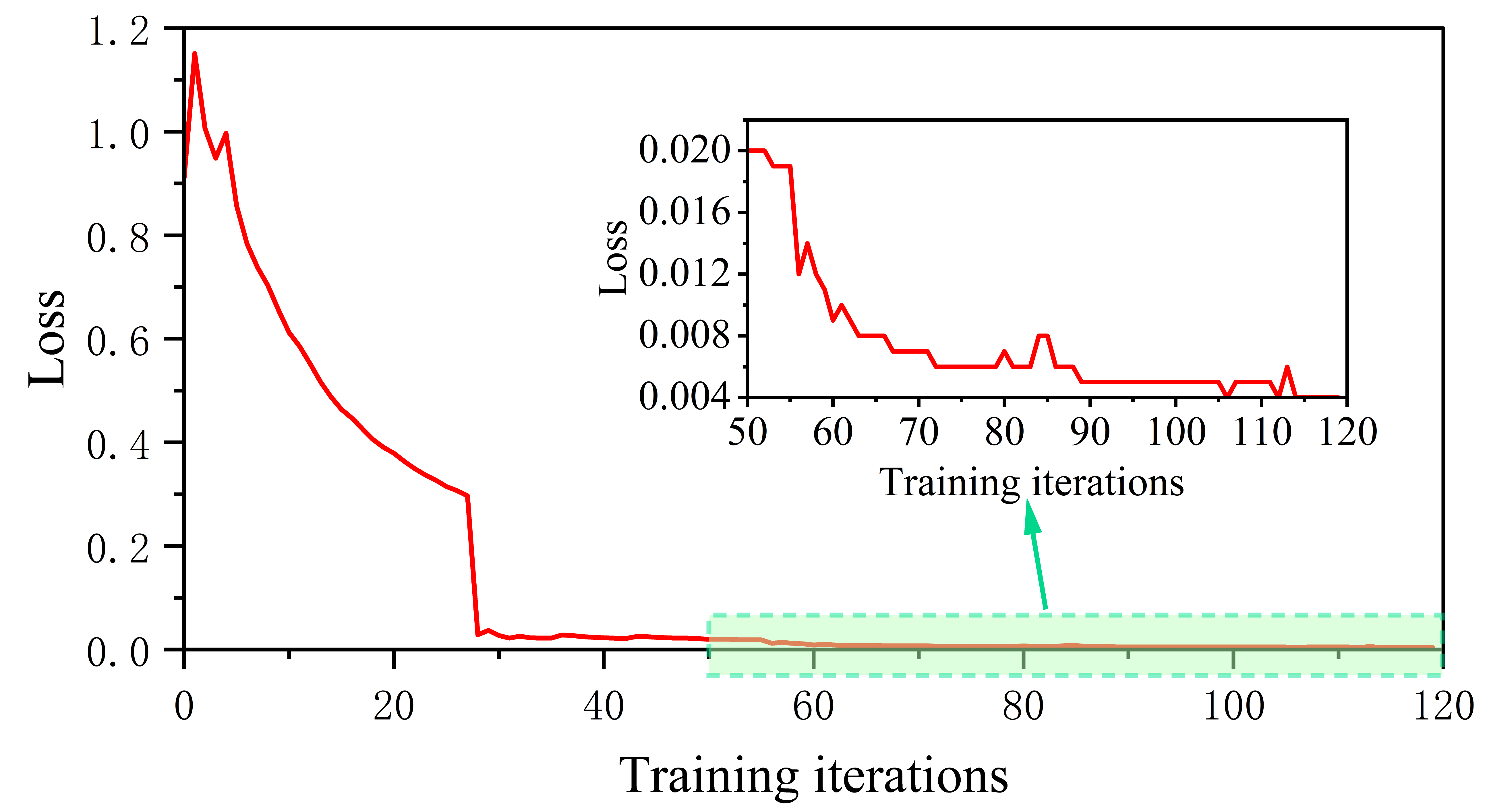}}\\
    \subfloat[]	{\includegraphics[width=\columnwidth]{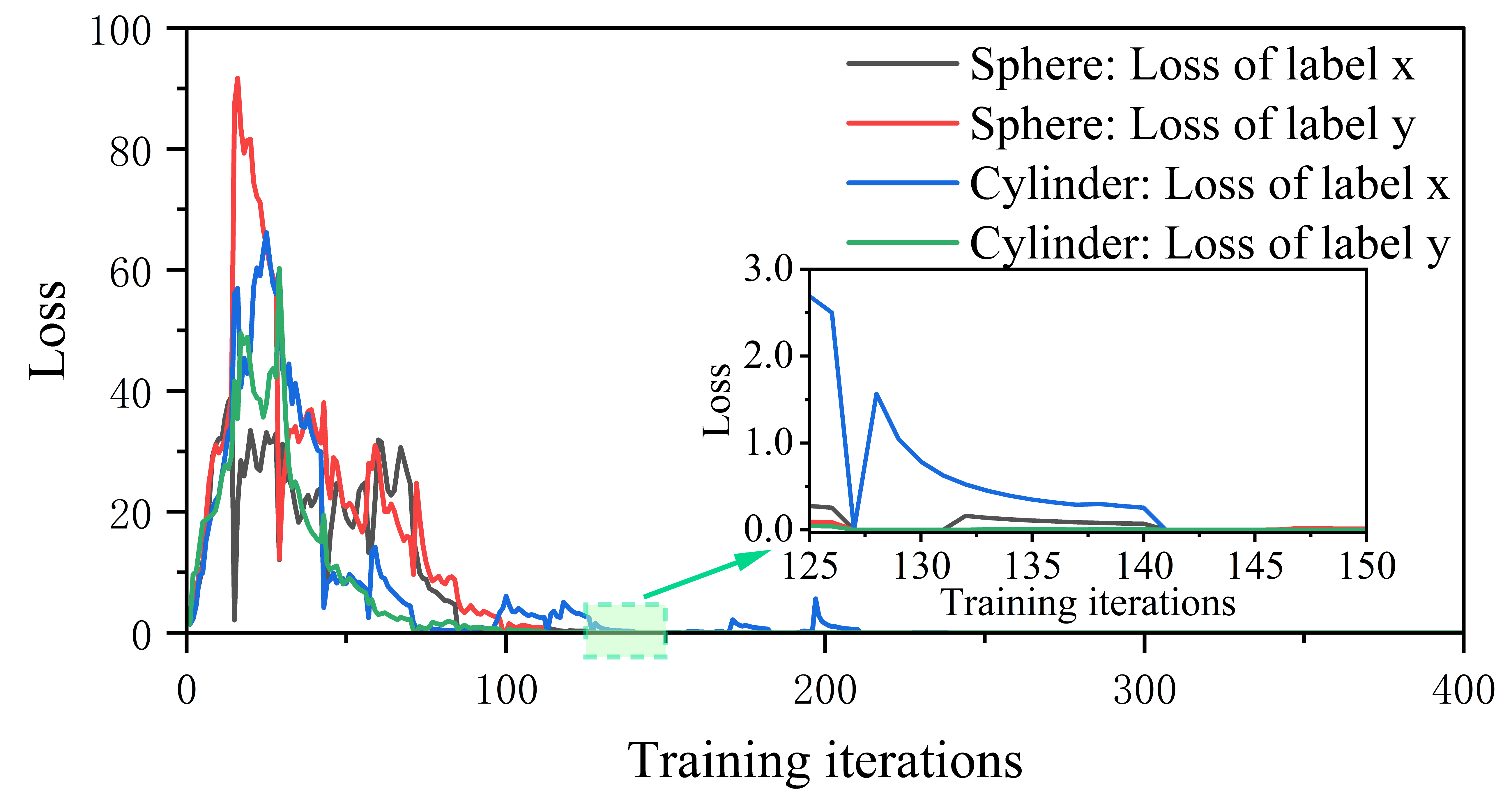}}
    \caption{Training loss of the RLCNN based on experimental data relative to training iterations: (a) recognition, (b) localization.}
    \label{Fig12}
\end{figure}

The experiments were conducted inside a metal cavity measuring 673mm${\times}$489mm${\times}$489mm, as depicted in Fig. \ref{Fig11}. Two monopole antennas, referred to as ${r_{1}}$ and ${r_{2}}$, operating within the frequency range of 5.3GHz-6.3GHz, are positioned on the inner walls of the metal cavity on the left and right sides respectively. To alleviate the effects of symmetry, the positioning grids are placed off-center at the bottom of the metal cavity, with each small grid having a side length of 36mm. The metal entity to be measured will be sequentially positioned within each small grid. To verify the effectiveness of the method in recognizing and locating subwavelength scatterers, the entities to be measured are a metal sphere with a diameter of 12mm and a metal cylinder with a diameter of 12mm and a height of 12mm, both of which are about 1/4 wavelength of 5.8GHz. The frequency and entities are set different from the simulated cases to verify the universality of the method.

In the experimental processes, we measured the frequency-domain transmission responses between two antennas by using vector network analyzer and converted them to the time-domain responses to use the TRTFPPs calculation method in Section \ref{sectionII}. Based on the method detailed in Section \ref{sectionII}, we placed the sphere and cylinder in 16 different positions respectively and repeated the measurement process 8 times for each case. In this way, each entity generated a total of 128 TRTFPPs.

For the neural network recognition, the RLCNN proposed in Section \ref{sectionII} was used to recognize the TRTFPPs of the sphere and cylinder. Specifically, the training dataset of shape recognition was composed of 112 TRTFPPs for spheres and 112 TRTFPPs for cylinders, randomly selected. The training dataset comprised a total of 224 TRTFPPs, with the remaining TRTFPPs allocated for the testing dataset. For the two localization sub-networks, 112 of the 128 TRTFPPs of each were used as the training dataset, and the remaining served as the testing dataset.

\begin{figure}[!t]
    \centering  
    \subfloat[]	{\includegraphics[width=0.6\columnwidth]{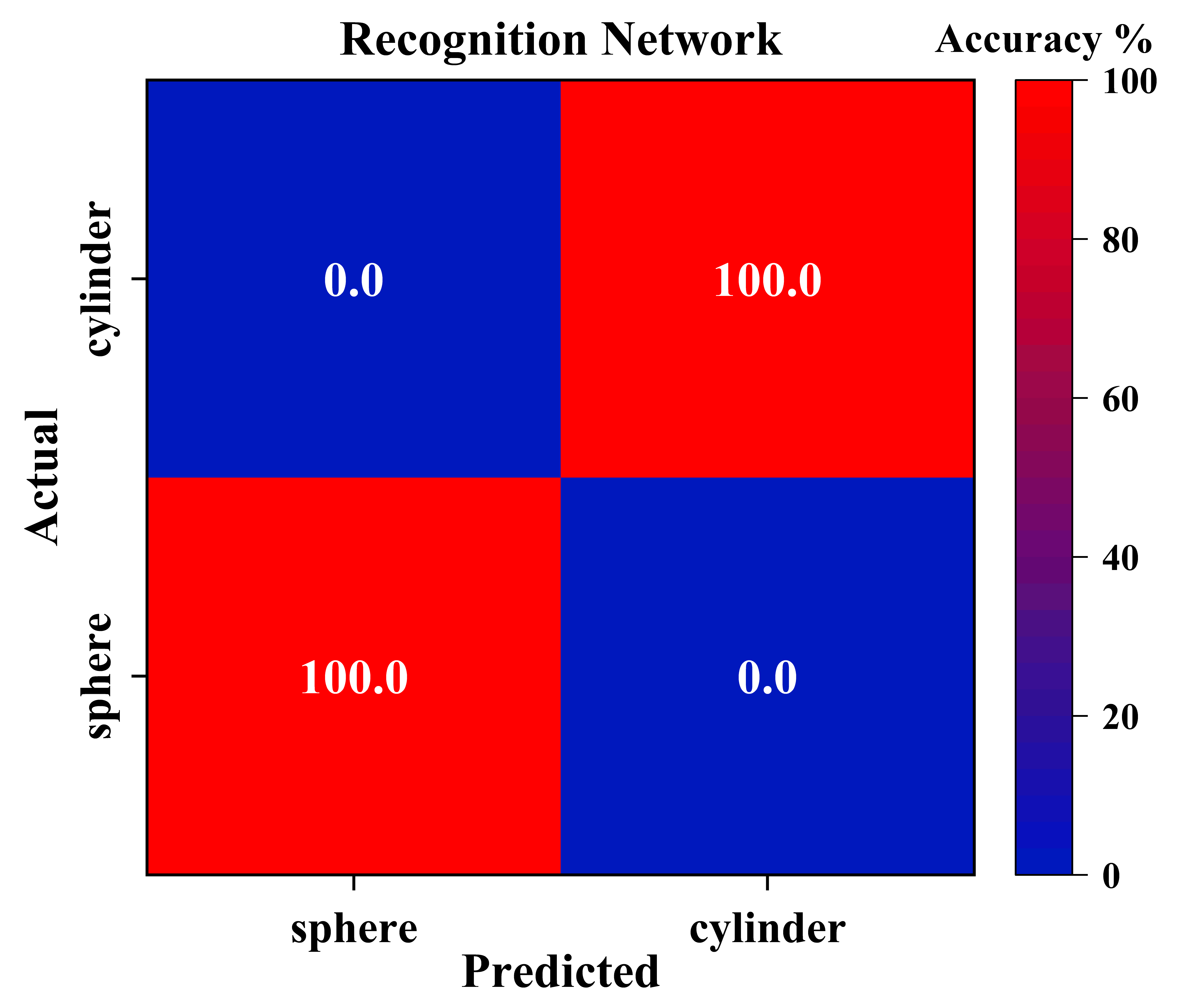}}\\
    \subfloat[]	{\includegraphics[width=\columnwidth]{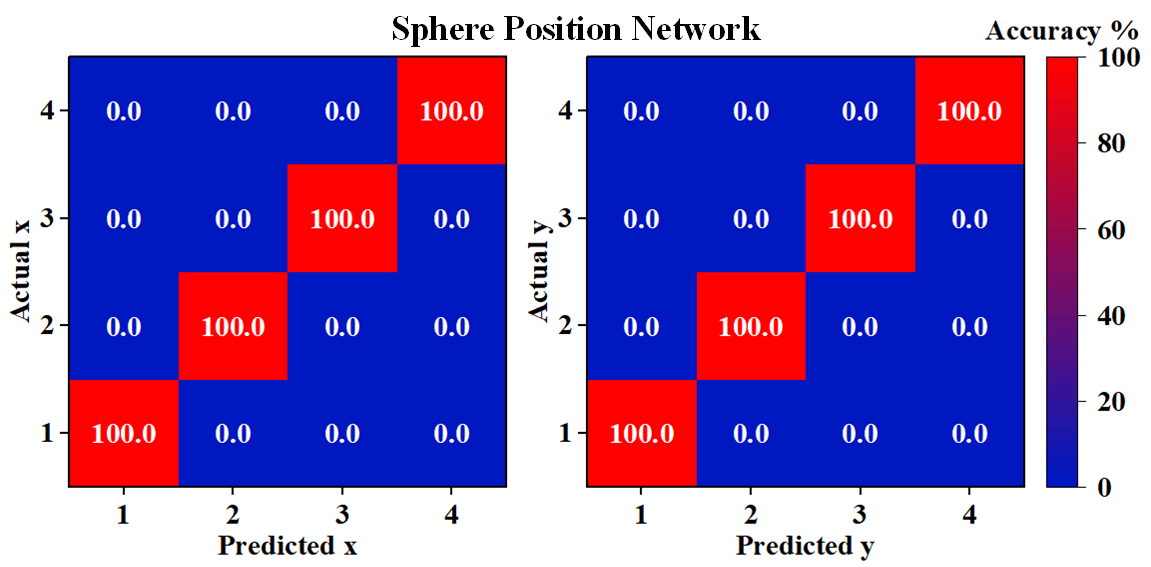}}\\
    \subfloat[]	{\includegraphics[width=\columnwidth]{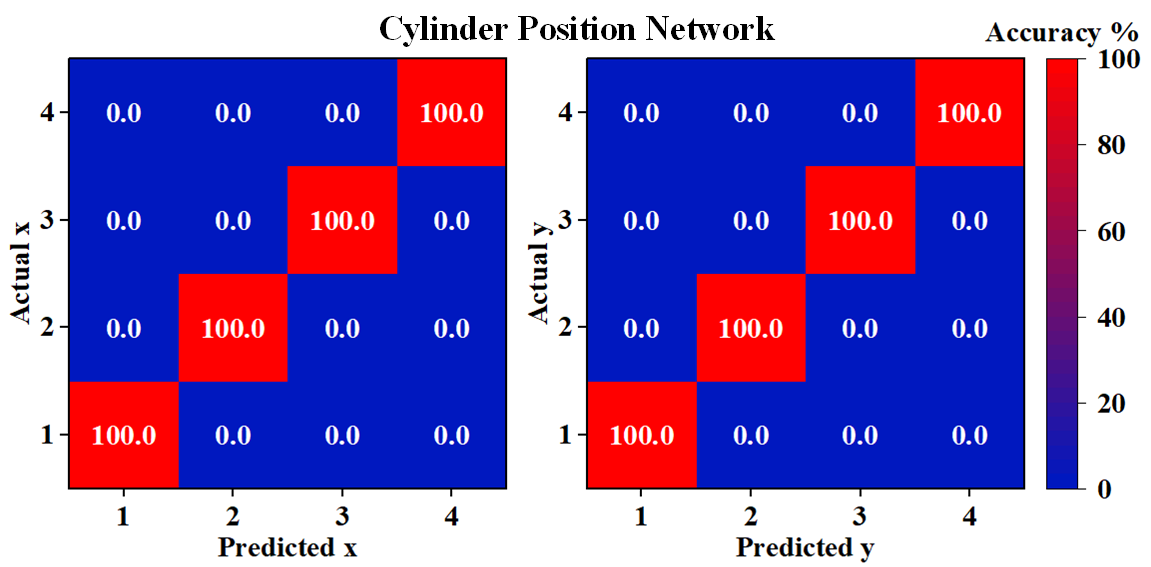}}
    \caption{Heat map of the RLCNN: (a) recognition, (b) sphere localization, and (c) cylinder localization. The horizontal label (x) and vertical label (y) in (b) and (c) indicate the 16 girds in Fig. \ref{Fig11}.}
    \label{Fig13}
\end{figure}

Figs. \ref{Fig12} and \ref{Fig13} present the training loss and heat map generated by the RLCNN based on experimental data, respectively. The rapid convergence of the training loss, attaining perfect 100\% accuracy in both recognition and localization, strongly validates the feasibility of our proposed method for practical applications.

Table \ref{Table3} offers a comparative analysis between the proposed technique and several established methods. It is evident that our method uniquely achieves concurrent recognition and localization of subwavelength non-cooperative entities through the utilization of a basic SISO system.
\begin{table}[!t]
\renewcommand\arraystretch{1.5}
\centering
\caption{Comparison of This Work with Other Methods}
\label{Table3}
\setlength{\tabcolsep}{3pt}
\begin{tabular}{m{60pt}<{\centering} m{30pt}<{\centering} m{20pt}<{\centering} m{20pt}<{\centering} m{20pt}<{\centering} m{20pt}<{\centering} m{30pt}<{\centering}}
    \hline
    & \cite{b7} & \cite{b11} & \cite{b13} & \cite{b15} & \cite{b19} & Our Work\\ 
    \hline
    System & E-pulse radar & SISO & MIMO radar & MA-SISO${^{*}}$ & SISO & SISO \\
    Subwavelength & No & No & No & Yes & No	& Yes\\
    Non-Cooperative & Yes & Yes & Yes & Yes & No & Yes\\
    Recognition & Yes & Yes & Yes & No & No & Yes\\
    Localization & No & Yes & Yes & Yes & Yes & Yes\\
    \hline
\multicolumn{7}{p{230pt}}{${^{*}}$ MA-SISO represents the metasurface- assisted single input single output (SISO) system.}\\
\end{tabular}
\end{table}

\section{Conclusion}
In this paper, we introduce a novel method that concurrently addresses the challenges of non-cooperative localization and recognition of subwavelength entities within complex multi-scattering scenarios. Through the fusion of TRTFPPs and neural networks, our investigation makes a significant stride in surmounting the intricacies tied to subwavelength non-cooperative entities within multi-scattering scenarios. This endeavor yields profound insights and opens avenues for applications spanning various domains that demand precise target localization and recognition.

Although the positions of the entities in this paper are represented by grid labels, it is noteworthy that the localization and recognition accuracy have reached the sub-wavelength level. By further subdividing the grid and establishing a more comprehensive database, or by augmenting the dataset through interpolation and other techniques, it is expected that even more precise real-time localization and recognition can be achieved. This will be the focal point of our subsequent research efforts.

\end{document}